# An Authentication Protocol Based on Combined RFID-Biometric System

Noureddine Chikouche
Department of Computer Science
University of M'sila
Algeria

Foudil Cherif
LESIA Laboratory
University of Biskra
Algeria

Mohamed Benmohammed
LIRE Laboratory
University of Constantine
Algeria

*Abstract* — **Radio Frequency Identification (RFID) and biometric technologies saw fast evolutions during the last years and which are used in several applications, such as access control. Among important characteristics in the RFID tags, we mention the limitation of resources (memory, energy, …). Our work focuses on the design of a RFID authentication protocol which uses biometric data and which confirms the secrecy, the authentication and the privacy. Our protocol requires a PRNG (*Pseud-Random Number Generator*), a robust hash function and Biometric hash function. The Biometric hash function is used to optimize and to protect biometric data. For Security analysis of protocol proposed, we will use AVISPA and SPAN tools to verify the authentication and the secrecy.**

## I. INTRODUCTION

At present, the problem of access control is very important in several applications. Physical access control consists in verifying if a person asking to reach a zone (e.g. building, office, parking, laboratory, etc.), has the right necessities to make it. The protocols of identity verification which allow access are called the authentication protocols. They answer the following two questions: "Who am I?" and "Am I really the person who is proceeding?". Answer to this first question is based on the recognition or the identification of the user which consists in associating an identity to a person, such as a smartcard or a RFID tag. Concerning the second question which articulates on the verification or the authentication of the user, it gives permission to a proclaimed identity. In other terms, it consists in identifying a user from one or several physiological characteristics (fingerprints, face, iris, etc.), or behavioural (signature, measure, etc.). These techniques are called Biometric Methods [14].

Among techniques and systems of identification which were developed quickly during the last years, we can notice that Radiofrequency identification (RFID) that is used in different domains (health, supply chain, access control, etc.). The RFID systems consist of three entities: (1) the tag (or the label) is a small electronic device, supplemented with an antenna that can transmit and receive data, (2) the reader which communicates with the tag by radio waves and (3) the server (or database, back-end) which uses information obtained from the reader for useful purposes. The main characteristic of a RFID system is the limitation of resources (memory, the processor, the consumption of energy, etc. …); on the other hand, these systems are necessary to assure security in all the levels of the system. Major difference between a RFID tag and a contactless smartcard is the limitation of computer resources.

In RFID systems, several authentication protocols have been developed [4,5,6,7]. Difference between these techniques lies in the realized properties of security and the complexity of implementation. Most of these protocols answer the first question only "Who am I? ". On the contrary systems with smartcards there are several authentication protocols based on the biometric technology, we mention here [8,9,10].

This paper, we propose an authentication protocol based on the combination between a RFID system and a biometric system. We verify secrecy, authentication of the tag and authentication of the reader by AVISPA&SPAN tools [1,2]. The conceived protocol protects the privacy of the user. To estimate these performances, we will compare it with the other RFID protocols and the biometric protocols of smart cards.

This paper is organized as follows: section II presents related work. Section III presents the system and hypotheses. Section IV presents the proposed protocol. The section V presents a check of the protocol automatically. An analysis of privacy side is then presented. Section VI presents a comparison of performances with existing works. We end by a general conclusion.

## II. RELATED WORK

In the protocols using identifier ID, two mechanisms are used: static and dynamic. The characteristic of the mechanism of static ID is that the identifier of the tag remains the same during the complete authentication, but that of the dynamic mechanism, the identifier of tag is modified. Every mechanism has his advantages and inconveniences. Here we present mainly the mechanism of static ID used in this article.

In the RHLS protocol (Randomized Hash Lock scheme) [5], information passed on with the tag every time when it is interrogated consists of random value nt and value H1 = h(ID, nt). RHLS which discovers two types of attacks: replay attack and tracing attack.

Concerning the protocol which is proposed with Chien and Huang (CH protocol) [4], the reader R and the tag T share secrets k and ID. Launch by the reader which sends a nonce nr. The tag produces an unpredictable nonce nt and calculates the hash function *g*, such as $g = h(nr \oplus Nt \oplus ID)$. This hash function and ID are used as parameters for the function rotate. The value of ID is returned; it depends on the value of g. The tag calculates the xor of the returned ID and g, before the sending of the left half of the results and nt to the reader. The reader calculates every pair of ID and k until it finds the corresponding tag. It sends then the right half of the xor of ID returned and g to the tag. CH protocol which discovers an attack of the type algebraic replay attacks, its cause is the false use of the algebraic operator xor in messages passed on by the function g. This attack is discovered also in LAK protocol [19].

Lee and al. [7] propose a protocol improved to avoid two types of attack: tracing and spoofing attack by use of various values of the hash function h during every authentication. These objectives are realized after analysis of this protocol. The cost of a hash function operation in tag is four, what is incompatible with the storage space and lower capacity to calculate. Therefore, excessive calculation will affect inevitably the efficiency of the protocol.

Biometry is widely used in the authentication protocols of the smart cards applications [8, 9, and 10]. The use of these protocols in RFID systems will depend on the availability of computer resources (memory, complexity, performance …) in the constituents of RFID systems and especially the RFID tag. The recent protocol [10] requires the calculation of seven operations of the function h in the phases of login and authentication and requires 4l as storage space in the tag. This number of calculations and this storage space influences negatively on the efficiency of a RFID protocol. Another difficulty concerns "Matching" treatment. In the biometric authentication protocols, this part is made in the smart card with the technique Match-on-card.

Concerning the material implementation of combined systems biometric-RFID, we shall quote two recent works. Rodrigues and al. [15] propose a decentralized authentication solution for embedded systems that combines both token-based and biometric-based mechanism authentication. Aboalsamh [16] propose a compact system that consists of a CMOS fingerprint sensor (FPC1011F1) is used with the FPC2020 power efficient fingerprint processor; which acts as a biometric sub-system with a direct interface to the sensor as well as to an external flash memory for storing finger print templates. An RFID circuit is integrated with the sensor and fingerprint processor to create an electronic identification card (e-ID card). The e-ID card will pre-store the fingerprint of the authorized user. The RFID circuit is enabled to transmit data and allow access to the user, when the card is used and the fingerprint authentication is successful.

III. SYSTEM AND HYPOTHESES

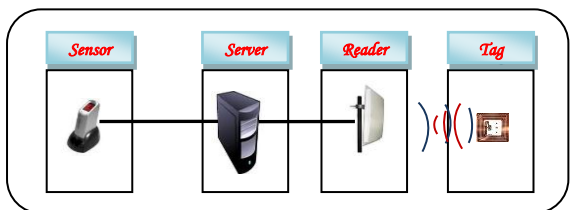

Figure 1. RFID-Biometric System

A. *System modeling*

The proposed system of authentication is based on the combination of two sub-systems: a RFID system and a biometric system. RFID system consists of: a tag ($\mathcal{T}$), a reader ($\mathcal{R}$) and a server ($\mathcal{S}$). Used biometric system consists of two entities, a sensor ($\mathcal{SR}$) and a server ($\mathcal{S}$), see the Figure 1.

*1) Biometric data*

Biometric data can be stored in the tag or in the data base. The biometric template will be stored in the tag. It offers a greater privacy and the mobility of the user. This assures also that information will always be with the user's tag. Storing the raw biometric data typically requires substantially more memory. For example, a complete fingerprint image will require 50 to 100 Kbytes, while a fingerprint template requires only 300 bytes to 2 Kbytes [14]. This condition is not always sufficient especially for the type of passive RFID tags. In our system, a practicable solution to optimize and to protect biometric data is the hash function. This function of template allows pressing the biometric template in an acceptable size.

The problem which lies with the hash functions standard (e.g. SHA-1, MD5, SHA-256, …) is comparison between two templates: the template which is protected in the tag a $h(B)$ and the template which is generated from the capture $h(B')$. Equality $h(B) = h(B')$ for the same person is not always assured, because $B'$ is a dynamic template where the person never keeps the same biometric features, (e.g. movement of the finger during the purchase), which implies the existence of a rate of error. We will quote two research works:

Sutcu and *al.* [12] propose a secure biometric based authentication scheme which fundamentally relies on the use of a robust hash function. The robust hash function is a one-way transformation tailored specifically for each user based on their biometrics. The function is designed as a sum of properly weighted and shifted Gaussian functions to ensure the security and privacy of biometric data. They also provide test results obtained by applying the proposed scheme to ORL face database by designating the biometrics as singular values of face images.

A. Nagar and al. [13] propose six different measures to evaluate the security strength of template transformation schemes. Based on these measures, they analyze the security of two well-known template transformation techniques, namely, Biohashing and cancelable fingerprint templates based on the proposed metrics.

*2) Tag and Reader*

The tag stores the identity (ID) and the biometric hash function of the template of the person (GB). This ID is strictly confidential and is shared between the database of the back-end server ($\mathcal{S}$) and the tag ($\mathcal{T}$). The tag can generate random numbers and calculation of the hash function *h* of a number. Standard ISO and EPC GEN2 (*Electronic Product Code, Generation 2*) support to produce the random numbers (nonces) in the tag. The reader $\mathcal{R}$ can generate also the random numbers.

*3) Server*

The server has two main functionalities:

- For the biometric system: extraction of the characteristics of a biometric modality to create a model or template (*B*),
- Concerning RFID system: it contains the database which includes the list of the identity of tags (*ID*).

*4) Sensor*

A biometric sensor is an electronic device used to capture a biometric modality of a person (fingerprint, face, voice, etc.).

### B. Security and privacy requirements:

Our protocol strives to achieve four requirements: secrecy, authentication of the tag, authentication of the reader and untraceability.

- Secrecy: or confidentiality, the verification that the identity of the tag ID is never passed on clearly to air on the interface radio frequency which can be spied.
- Authentication of tag: A reader has to be capable of verifying a correct tag to authenticate and to identify a tag in complete safety.
- Authentication of reader: A tag has to be capable of confirming that it communicates with the legitimate reader (a single reader exists in communications between the constituents of the RFID system).
- Untraceability: We consider the notion of untraceability as defined in [17] which captures the intuitive notion that a tag is untraceable if an adversary cannot tell whether he has seen the same tag twice or two different tags.

### C. Intruder Model

Besides modelling security protocols, it is also necessary to model the intruder, that is to say, to define its behaviour and limit. For this, the assumptions used are gathered under the name "Dolev-Yao model" [6]. This intruder model is based on two important assumptions that are *the perfect encryption* and the *intruder is the network*.

Perfect encryption ensures in particular that an intruder can decrypt a message *m* encrypted with key *k* if it has the opposite of that key. The second hypothesis which is "*the intruder is the network*" means that, the intruder has complete control over the network and he can derive new messages from his initial knowledge and the messages received from honest principals during protocol runs. To derive a new message, the intruder can compose and decompose, encrypt and decrypt messages, in case he knows the key.

For the assumption "*the intruder is the network*", the RFID network system in this case is wireless, it is based on communication by radio waves. Communication among the server and the reader and between the server and the sensor is secure. Contrary to this, communication between the tag and reader is not assured and based on radio frequencies waves. We assume that the adversary can observe, block, modify, and inject messages in any communication between a reader and a tag.

### IV. PROPOSED PROTOCOL

The proposed Protocol is divided into two processes: the phase of registration and the phase of mutual authentication. We, afterward, use the following notations:

| | |
|---|---|
| T | RFID tag or transponder |
| R | RFID reader or transceiver |
| S | Back-end server |
| Nt | random number (nonce) generated by tag T |
| Nr | random number (nonce) generated by reader R |
| H() | One-way hach function |
| G() | BioHash (Biometric hash function ) |
| ‖ | Concatenation of two inputs |
| B | Biometric template |
| ID | Identity of tag |
| GB | Biohashed value of B |
| ⊕ | Or-exclusif |
| $H_R$ | The right half of H |
| $H_L$ | The left half of H |
| X≈Y | mean X=Y±E  (E : rate of error) |

Steps detailed by two processes are described below.

### A. Registration Processus

This initial phase called also registration. The objective is to create a template biometric and stored in related to a declared identity (see the figure 2.). In this phase, it has to execute the following steps to obtain the RFID tag.

*Step 1*: the authorized user inputs his/ her personal biometrics, to pass it on to the server of the trusted registration center (*RC*).

*Step 2*: the *RC*, after extraction of biometric characteristics, creates a biometric template B, and computes the biometric hash function GB such as GB = g (B).

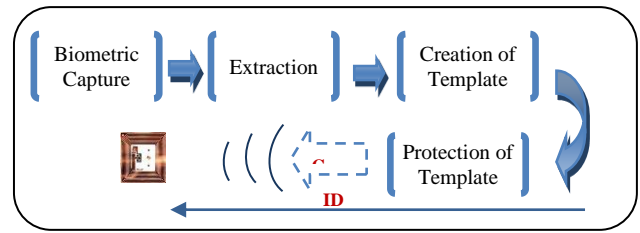

Figure 2.  Registration Processus

*Step 3*: Then, the registration center stores the information {ID, GB} in the user's tag and sends it to the tag through a secure channel.

$$\mathcal{RC} \xrightarrow{ID, G\mathrm{B}} \mathcal{T}$$

### B. Mutual Authentication Processus

According to the order of the passed on messages, the process of authentication takes place as follows

(to see Figure 3):

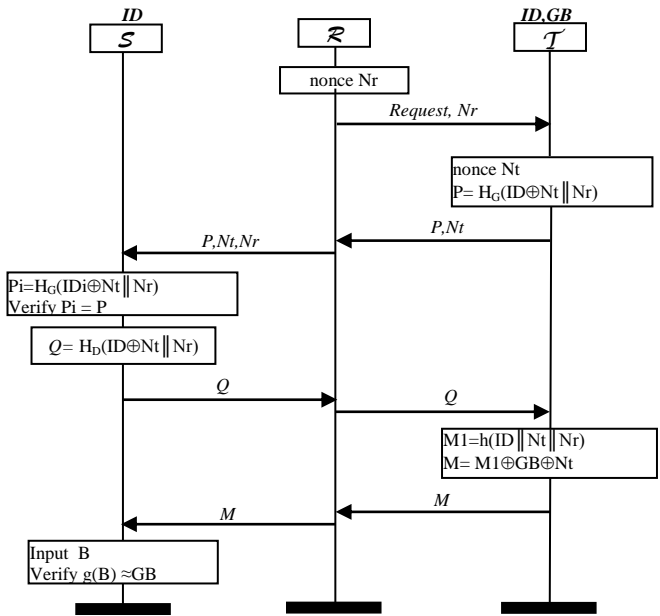

Figure 3. Proposed Protocol

*Step 1: Challenge*

The reader RFID produces a nonce Nr and sends it then, and a request to the tag. Three cases can occur: *1)* No tag answers, *2)* A tag answers, *3)* Many tags answer at the same time. In our protocol, the last case is not approved because one requires the capture of only one biometry for every person and for every tag in every process of authentication.

*Step 2: Authentication of the tag*

*Step 2.1:* the tag found in the step 1 generates a nonce Nt and computes P = $H_G$(ID⊕Nt∥Nr)

*Step 2.2:* the tag sends P with the nonce Nt to the reader RFID,

*Step 2.3:* the reader resends the successful P message, Nt and the nonce Nr to the server.

*Step 2.4:* from the database, the server looks for certain IDi (such as $1 \leq i \leq n$, n the number of tags) to compute Pi=$H_G$(IDi⊕Nt∥Nr), and make the following comparison:

$$Pi \;?=\; P$$

If it is found, the tag crosses the authentication of the tag and is considered as legitimate, otherwise to end.

*Step 3: Authentication of the server*

*Step 3.1*: the server computes and sends to the reader Q;

$$Q = H_D(ID_i \oplus Nt \| Nr) \text{ such as } ID_i = ID$$

*Step 3.2*: the reader sends the Q message in the tag.

*Step 3.3*: the tag computes $H_D$(ID⊕Nt∥Nr) and verifies if:

$$Q \;?=\; H_D(ID \oplus Nt \| Nr)$$

If they are equal, the authentication of the reader is successful; otherwise the authentication of the reader has failed.

*Step 4: Verification of biometry*

*Step 4.1*: the tag computes M1 = h (ID∥Nt∥Nr) and makes operation or-exclusive of M1 with GB and Nt. The resultant message is M = M2⊕GB⊕Nt.

*Step 4.2*: the tag sends M to the reader RFID, and the reader resends received message to the server.

*Step 4.3*: after acquiring of the biometry of the user from the sensor, it sends it to the server. The server extracts biometric characteristics and generates the template B. the server computes the biometric hash function of the template g(B).

*Step 4.4*: from the database, the server computes M2=h (IDi∥Nt∥Nr)⊕Nt, such as IDi= ID (of the step 2.4), and extracts GB from:

$$M2 \oplus GB = M$$

*Step 4.5*: to make the comparison of type 1:1 of g(B) ≈ GB, if it is confirmed, the person is a trusted user, otherwise, the bearer of the tag is illegitimate, the information of failure will be sent to the reader, the protocol is interrupted.

## V. SECURITY VERIFICATION OF PROTOCOL

### A. Automatic Verification

There are several tools of automatic verification of protocols. We chose tools AVISPA (*Automated Validation of internet Security Protocols and Applications*) [1] and SPAN (*Security Protocol ANimator*) [2] for the following reasons: four tools are available using various techniques of validation (Model-checking, automate trees, resolution of constraints, Solver SAT). These tools are based on the same language of specification: language HLPSL (*High-Level Protocol Specification Language*) [18]. The platform AVISPA is the analyzer which models a big number of protocols (more than 84 protocols). Among these four tools, two tools OFMC and CL-ATSE which can verify protocols requiring the operator or-exclusive (xor). Concerning our protocol, we verify the properties of the confidentiality of the identity ID (sec_id_TR and sec_id_RT respectively), the confidentiality of the template B ( sec_b ), the authentication of the tag (aut_tag) and the authentication of the reader ( aut_reader). These properties are specified in HLPSL as follows:

```
goal
    secrecy_of sec_b, sec_id_TR, sec_id_RT
    authentication_on aut_reader
    authentication_on aut_tag
end goal
```

Concerning the authentication, there are two possible attacks: the replay attack and the attack Man-in-the-Middle. For it, we uses two types of specification in the role HLPSL's environment.

*1) Replay Attack*

In the replay attack, the intruder can listen to the message of answer of the tag and to the reader. It will broadcast the message listened without modification to the reader later.

Specification below of the role environment in HLPSL depends on the treatment of two identical sessions between the same tag and the same reader ($t$ and $r$). This scenario allows discovering the attacks of the type replay attack if it exists.

```
role environment() def=
const t,r : agent,
      id,b : text,
      h,g,left,right : hash_func
intruder_knowledge = {t,r,h,g,hright,hleft}
composition
session(t,r,id,b,h,g,hright,hleft) /\
session(t,r,id,b,h,g,hright,hleft)
end role
```

After the verification of this protocol by AVISPA tools, result is as follows:

```
SUMMARY
  SAFE
DETAILS
  BOUNDED_NUMBER_OF_SESSIONS
  UNTYPED_MODEL
PROTOCOL
  C:\progra~1\SPAN\testsuite\results\BioMRFID.if
GOAL
  As Specified
BACKEND
  CL-AtSe
STATISTICS
  Analysed   : 600 states
  Reachable  : 188 states
  Translation: 0.01 seconds
  Computation: 0.02 seconds
```

This result means in clearly that there is no replay attack. We can thus deduct that the diagnosis of AVISPA&SPAN tools for this protocol is secure.

*2) Main-in-the-midlle Attack*

The scenario of the role `environment` below allows discovering the attacks of this type if it exists.

```
role environment() def=
const t,r : agent,
      id,b,idti,idri,bti,bri : text,
      h,g,hright,hleft : hash_func
intruder_knowledge = {t,r,h,g,
hright,hleft,idti,idri,bti,bri}
composition
    session(t,r,id,b,h,g,hright,hleft)
 /\ session(t,i,idti,bti,h,g,hright,hleft)
 /\ session(i,r,idri,bri,h,g,hright,hleft)
end role
```

The result of the check with this scenario is the same as with the scenario a). We can thus deduct that this protocol is resistant in the attack of the "man in the middle".

*B. Security Analysis*

We now analyze the security properties of the proposed protocol as follows: untraceability, desynchronization resistance and with Denial of service (DOS) attack prevention.

*1) Untraceability :*

During every session of authentication, an opponent can observe only the values of (Nt, Nr, M1, P, Q), where, Nt and Nr are random numbers and M1 and Q messages are calculated the right/ left part of the function H(ID⊕Nt‖Nr). The P message = H(ID‖Nt‖Nr)⊕GB⊕Nt. The opponent cannot deduce the value of ID because function H(ID‖Nt‖Nr) is very effective as is shown in the paper of [11]. In M1 messages, P and Q, the opponent cannot correlate ID and B because these two values are secret and Nt and Nr are random numbers changed in every authentication. So, an opponent cannot track tags.

*2) Desynchronization Resistance :*

Our protocol belongs to the static mechanism ID where the identifier of the tag is fixed. So, in the case of the loss of message, failing of energy or the loss of connection with the server during the authentication, it will not affect the database of the server and will not become an obstacle to the protocol.

*3) DOS attack Prevention:*

There are several categories of Dos attacks, one is to desynchronize the internal states of two entities, and the other is to exhaust the resources of the parties involved. For RFID authentication protocols, researchers are concerned about desynchronization.

For our protocol, the internal state ID is kept static and not changed during authentication process. So, it can resist the attack of denial of service.

In the Table I below, a comparison of the security with protocols mentioned early is given [4, 5, 6, and 7].

TABLE I : ANALYSIS OF SECURITY

| RFID Protocol (static ID) | RLHS [5] | LCAP [6] | CH [4] | LHYC [7] | Our Protocol |
|---|---|---|---|---|---|
| Mutual Authentification | + | + | + | + | + |
| Replay attack prevention | - | + | - | + | + |
| Non traceability | - | + | + | + | + |
| DoS attack prevention | - | - | + | + | + |
| Desynchronization Resistance | + | + | + | + | + |

## VI. PERFORMANCE ANALYSIS

As compared with what follows in Table II. This table illustrates the storage cost, the communication cost, and the computation cost of entities. The computation cost is a function of the number of operations of the hash function in phase's login and the authentication on the smartcard for the biometric protocols, as well as of the number of operations of the hash function on the tag in RFID protocols.

*Computation Cost:* the tag used in the protocol proposed by Lee and al. (Protocol LHYC) [7] and the smart cards of the biometric protocols require an important number of operations for the hash function. On the contrary, in the protocol of Chien and Huang [4], it requires a random numbers generator with an input number, but it is necessary not to forget the replay algebraic attack.

In our protocol, we require two operations of calculation of function h in the tag, so these calculations are effective for RFID tags.

*Communication Cost:* Communication cost between a tag and a reader consists of: the number of message exchanges, and the total bit size of the transmitted messages, per each communication. Concerning our protocol, the total of the bits of the messages of communication tag to the reader is: 2½l and for the message of communication reader to tag is: ½l. With regard to the other protocols of smart cards the performance of the communication of our protocol is more effective.

TABLE II : PERFORMANCE ANALYSIS

| Protocol | | Computation Cost Tag/Card | Storage Cost | Communication Cost | | |
|---|---|---|---|---|---|---|
| | | | | R→T | T→R | Σ |
| RFID | [4] | 1g | 2l | ½l | 1½l | 2l |
| | [5] | 1h | 1l | - | 2l | 2l |
| | [6] | 2h | 1l | 1l | 2l | 3l |
| | [7] | 4h | 2l | 1l | 2l | 3l |
| Smart Card | [8] | 4h | 3l | 2l | 3l | 5l |
| | [9] | 4h | 3l | 2l | 3l | 5l |
| | [10] | 3h | 4l | 2l | 3l | 5l |
| Our protocol | | 2h | 2l | ½l | 2½l | 3l |

Notations: *h* - the cost of a hash function operation,
 *g* - random number generator with an input number,
 *l*: size of required memory.

*Storage Cost:* The amount of storage needed on the back-end server is also another important issue. In the biometric protocols [8, 9], the smart card requires 3l bit and 4l for the protocol [10]. In our protocol, the tag requires 2l bit to store the identity (ID) and the function h of template (GB). Consequently, in the implemented protocols, the tag requires only 2l bits at most of the memory, which is adapted to tags with weak cost.

We can conclude that our protocol is effective and adapted to RFID tags as far as the computation cost; the storage cost and the communication cost are concerned.

## VII. CONCLUSION

We proposed in this article a new RFID authentication protocol which uses biometric data. Our protocol is compatible with the constrained computational and memory resources of the RFID tags. Concerning the problem of the size of biometric data, we applied the hash function to the biometric template, which allows to optimize and to protect these data. Our protocol realizes the secrecy private data, the authentication of the tag and the authentication of the reader. Experimental tests (with AVISPA and SPAN tools) proved it. We made an analysis of security on the efficiency of our protocol for untraceability, resistance for the denial of service (DOS) attack prevention and the desynchronization resistance.

The advantage of our protocol is that it can be used in decentralized applications since we have no need of biometric database of the users in the system.

Future research includes additional work in regards to the biometric hash function. There are many researches on the implementation of the robust hash function in RFID tags. But researches on the implementation of Biometric hash function are limited.